\documentclass[11pt]{article}

\usepackage[T1]{fontenc}
\usepackage[utf8]{inputenc}

\usepackage[margin = 0.9in]{geometry}
\usepackage{booktabs} 
\usepackage{enumitem}

\usepackage{latexsym,amsmath,amsfonts,amssymb,stmaryrd,mathtools}
\usepackage{amsthm}
\usepackage{xcolor}
\usepackage{algorithm, algorithmicx}
\usepackage[noend]{algpseudocode}
\usepackage{graphicx}
\usepackage{hyperref}
\usepackage{wrapfig}
\usepackage{listings, fancyvrb, multicol}
\usepackage{multirow}
\usepackage{times}
\usepackage{comment}
\usepackage{authblk}
\usepackage{xspace}
\usepackage{pifont}
\usepackage{parcolumns}
\usepackage{color,soul}

\usepackage{makecell}

\definecolor{ao}{rgb}{0.0, 0.5, 0.0}

\newcommand{\delayed}{delayed}
\newcommand{\protectedx}{protected}
\newcommand{\protects}{protects}
\newcommand{\protectx}{protect}
\newcommand{\protection}{protection}

\newtheorem{theorem}{Theorem}[section]

\newtheorem{definition}[theorem]{Definition}

\newcommand{\hide}[1]{}
\newcommand{\var}[1]{\textit{#1}}

\newcommand{\cfont}[1]{\mbox{\tt\bf\small #1}}
\newcommand{\typ}{\tau}
\newcommand{\emp}[1]{\epsilon_{#1}}
\newcommand{\cpy}{\cfont{copy}}
\newcommand{\destr}{\cfont{destruct}}

\newcommand{\newr}{\cfont{new}}
\newcommand{\loa}{\cfont{load}}
\newcommand{\str}{\cfont{store}}
\newcommand{\mov}{\cfont{move}}
\newcommand{\del}{\cfont{delete}}

\newcommand{\stdshared}{std::shared\_ptr}
\newcommand{\weakatomicstd}{Weak Atomic std::shared\_ptr}
\newcommand{\weakatomiccustom}{Weak Atomic Custom shared\_ptr}
\newcommand{\justthread}{just::thread}

\newcommand{\atomicsharedptr}{\cfont{atomic\_shared\_ptr}}

\newcommand{\swcopy}{\texttt{swcopy}}

\newcommand{\acquire}{\cfont{acquire}}
\newcommand{\release}{\cfont{release}}
\newcommand{\retire}{\cfont{retire}}
\newcommand{\ejectx}{\cfont{eject}}
\newcommand{\ejectall}{\cfont{ejectAll}}
\newcommand{\acqret}{Acquire-Retire}

\makeatletter
\lst@Key{countblanklines}{true}[t]%
{\lstKV@SetIf{#1}\lst@ifcountblanklines}

\lst@AddToHook{OnEmptyLine}{%
	\lst@ifnumberblanklines\else%
	\lst@ifcountblanklines\else%
	\advance\c@lstnumber-\@ne\relax%
	\fi%
	\fi}
\makeatother

\usepackage{subcaption} 
\begin{document}

\title{Concurrent Reference Counting and Resource Management in Constant Time}
  
  \author[1]{Guy E. Blelloch}
  \author[1]{Yuanhao Wei}
  \affil[1]{Carnegie Mellon University}
  \affil[ ]{\textit{\{guyb, yuanhao1\}@cs.cmu.edu}}
  \date{}

\maketitle

\begin{abstract}
  
A common problem when implementing concurrent programs is efficiently protecting against unsafe races between processes reading and then using a resource (e.g., memory blocks, file descriptors, or network connections) and other processes that are concurrently overwriting and then destructing the same resource.  Such read-destruct races can be protected with locks, or with lock-free solutions such as hazard-pointers or read-copy-update (RCU).

In this paper we describe a method for protecting read-destruct races with expected constant time overhead, $O(P^2)$ space and $O(P^2)$ delayed destructs, and with just single word atomic memory operations (reads, writes, and CAS).  It is based on an interface with four primitives, an acquire-release pair to protect accesses, and a retire-eject pair to delay the destruct until it is safe.  We refer to this as the acquire-retire interface.
Using the acquire-retire interface, we develop simple implementations for three common use cases:
(1) memory reclamation with applications to stacks and queues,
(2) reference counted objects, and
(3) objects manage by ownership with moves, copies, and destructs.
The first two results significantly improve on previous results, and the third application is original.
Importantly, all operations have expected constant time overhead.

\end{abstract}
\newcounter{results}
\newcommand{\result}[2]{\refstepcounter{results} \vspace{.03in}{\flushleft\textbf{Result \theresults~(#1)}: \emph{#2}}
\vspace{.05in}}

\section{Introduction}

In this paper, we are interested in the management of concurrently shared resources, and in particular in safely destructing them.  Resources can include memory, file pointers, communication channels, unique labels, operating system resources, or any resource that needs to be constructed and destructed, and can be accessed concurrently by many threads.  The problem being addressed is that when a handle to a resource is stored in a shared location, there can be an unsafe race between a thread that reads the handle (often a pointer) and then accesses its content, and a thread that clears the location and then destructs the content.  Such \emph{read-destruct} races are a common issue in the design of concurrent software.  In the context of allocating and freeing memory blocks, the problem has been referred to as the memory reclamation problem~\cite{michael2004hazard,brown2015reclaim} , the repeat offender problem~\cite{Herlihy05}, and read-reclaim races~\cite{hart2007performance}.  Elsewhere, read-destruct races come up in managing reference counts~\cite{just19asp}, OS resources~\cite{mckenney2013rcu}, and multiversioning~\cite{ben2019vm}.

Read-destruct races can be addressed by placing a lock around the reads and destructs, but this comes with all the standard issues with locks.  To avoid locks, many lock-free methods have been suggested and are widely used, including read copy update (RCU)~\cite{mckenney2008rcu}, epochs~\cite{fraser2004practical}, hazard-pointers~\cite{michael2004hazard}, pass-the-buck~\cite{Herlihy05}, interval-based reclamation~\cite{wen2018interval}, and others~\cite{ramalhete2017brief, brown2015reclaim}.  RCU is implemented in the Linux operating system, and McKenney et. al. point out that it is used in over 6500 places, largely to protect against read-destruct races~\cite{mckenney2013rcu}.  Epoch-based reclamation and hazard pointers are also widely used for memory management in concurrent data structure~\cite{david2015asynchronized}~\cite[Chapter~7.2]{williams2012book} to protect against read-destruct races.

Our contribution is to solve the read-destruct problem in a constant time (expected) wait-free manner---that is to say that both the read sections and the destructs can be implemented safely with expected constant time overhead.  Furthermore, our solution is memory efficient, only requires single-word CAS, and ensures timely destruction.
Our solution is expected rather than worst-case constant time because each process uses a local hash table.

\begin{figure}
\centering
    \includegraphics[scale=0.20]{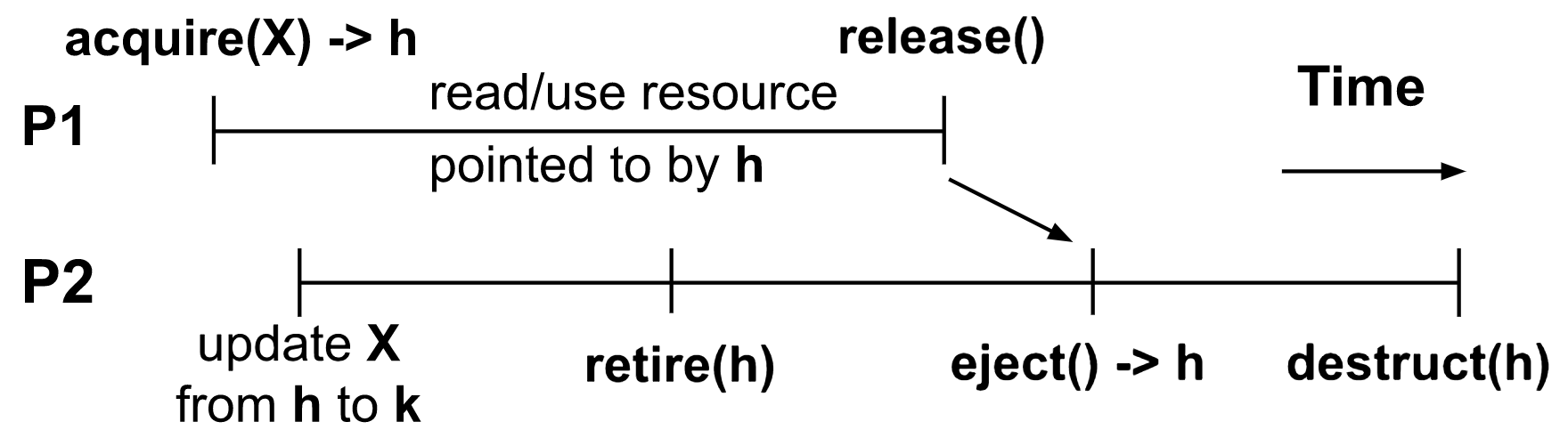}
    \caption{Using \acqret{} to protect against read-destruct races.}
    \label{fig:acqret}
\end{figure}

Specifically, we support a \emph{acquire-retire} interface consisting of four operations: acquire, release, retire, and eject.  An \emph{acquire} takes a pointer to a location containing a resource handle, reads the handle and \emph{\protects} the resource, returning the handle.  A later paired \emph{release}, releases the \protection.  A \emph{retire} follows an update on a location containing a resources handle, and is applied to the handle that is overwritten to indicate the resource is no longer needed.  A later paired \emph{eject} will return the resource handle indicating it is now safe to destruct the resource (i.e. it is no longer \protectedx{}). This definition is illustrated in Figure \ref{fig:acqret}.  We allow the same handle to be retired multiple times concurrently, each paired with its own eject.  This is useful in our reference counting collector where the destruct is a decrement of the reference count.
All operations are linearizable~\cite{herlihy1990linearizability}, i.e., must appear to be atomic.
Section~\ref{sec:related} compares the acquire-retire interface to similar interfaces.

As an example, Figure~\ref{fig:stream} defines a structure for protecting a resource based on acquire-retire, and uses it for redirecting a file stream.  The \cfont{use} function wraps a function call (on $f$) in an acquire-retire, while the \cfont{redirect} overwrites the value of the resource.  A thread (\cfont{other}) uses the stream concurrently while the main thread redirects the stream.  Without acquire-retire, there is a read-destruct race that would allow \cfont{other} to write to a stream that has already been closed.  Here, the \cfont{retire} binds a function (the destructor) that is applied when ejected.

\begin{figure}
  \small \center
\textbf{Protecting a Resource}
\begin{lstlisting}[linewidth=.99\columnwidth]
template <|typename R|>
struct protect {
  // construct protected resource from ptr
  protect(R* ptr) : p(ptr) {}

  // safely use the resource in function f
  template<|typename T, typename F|>
  T use(F f) {
    T result = f(acquire(&p));
    release();
    return result; }

  // redirect to a new resource, destructing old
  void redirect(R* newr) {
    R* oldr = fetch_and_store(&p, newr);
    retire(oldr);
    optional<|R*|> e = eject();
    if (e.has_value()) delete e.value();}
  ...
private: R* p;};
\end{lstlisting}
  \textbf{Example Use}
\begin{lstlisting}[linewidth=.99\columnwidth]
protect<|ofstream|> outstr(new ofstream("log/monday"));

// fork thread and use stream
thread other ([&] () {
   outstr.use<|void|>([] (ostream* str) {
     *str << "hello world" << endl;});});

// redirect is concurrent with use
outstr.redirect(new ofstream("log/tuesday"));
\end{lstlisting}
  \caption{A structure for protecting a resource from read-destruct races based on acquire-retire,
    and an example use for an output stream with a protected race.  Code in C++.
}
   \label{fig:stream}
\end{figure}

The acquire-retire interface can be implemented using hazard pointers~\cite{michael2004hazard} or pass-the-buck~\cite{Herlihy05}.     With these approaches, however, the acquire is only lock-free.   In particular after installing
the value read from the handle location into a hazard slot, it is necessary to go back and check that the location has not been
updated, and repeat if it has.
Epoch-based reclamation can be used to support the operations in constant time, it provides no bound on the number of delayed retires. We say a retire-eject pair is \emph{\delayed{}} between the retire and the eject.
 


In this paper we describe an efficient implementation of acquire-retire, which gives the following results.

\result{Acquire-Retire}{
\label{result:acqret}
  For an arbitrary number of resources and locations, $P$ processes, and at most $c$ resources \protectedx{} on each process at a given time, the acquire-retire interface can be supported using:
\begin{enumerate}
\item $O(1)$ time for acquire, release, and retire,
\item $O(1)$ expected time for eject,
\item $O(cP^2)$ space overhead and $O(cP^2)$ \delayed{} retires, and
\item single-word (at least pointer-width) read, write and CAS.
\end{enumerate}
}

For the space and \delayed{} resources bounds, we assume that every retire comes with at least one eject.  We note that $O(1)$ time for each operation is stronger than wait-free since wait-free only guarantees finite time.
Also, for $N$ resources and $P \in o(\sqrt{N})$, the space overhead is a low order term compared to the space for the resources themselves.  Limiting ourselves to pointer-width words implies that we do not use unbounded counters.
Such counters are used in many other non-blocking data structures to avoid the ABA problem~\cite{jayanti2005many,michael2004aba}.  We know of no previous solution that is constant (expected or worst-case) time, bounded space, and does not use much more powerful primitives.
Our results make use of recent results on atomic copy~\cite{BW19}.



\newcommand{\imr}{protected-block memory reclamation}
\paragraph{Applications of Acquire-Retire.}

We describe several applications that can use acquire-retire, and significantly improve on previous bounds for each of these.  We first consider the classic memory reclamation problem solved by hazard pointers~\cite{michael2004hazard} and pass-the-buck~\cite{Herlihy05}.  Here, the resources are memory blocks managed with explicit allocates and frees.  A read-destruct race can happen when one process reads a variable with a pointer to a block and uses the block, and a concurrent process overwrites the variable and frees the overwritten block.  We say the \emph{\imr}\footnote{We call it ``protected-block'' to distinguish from other ``protected-region'' interfaces such as epochs or RCU, which protect all blocks in a region of  code instead of individual blocks.} problem is to support
a \cfont{protected\_read} that reads and uses a pointer from a location, and a \cfont{safe\_free} that is applied when
the last pointer to a block is overwritten and ready to free, but only actually freed when it is not  \protectedx.
We say a block is \emph{\delayed{}} if it has been safe freed
but has not been actually freed.
We describe an implementation with the following bounds.  
 
\result{Memory Reclamation}{
  \label{result:reclamation}
For arbitrary number of blocks and $P$ processes each with at most $c$ \protectedx{} blocks at any time, 
 \imr{} can be supported with:
 \begin{enumerate}
\item $O(1)$ time overhead for \cfont{protected\_read}
\item $O(1)$ expected time for \cfont{safe\_free}, 
\item $O(cP^2)$ space,
\item $O(cP^2)$ \delayed{} blocks at any time, and
\item single-word (at least pointer-width) read, write, and CAS.
\end{enumerate}
}

Hazard pointers solve the problem, but with unbounded time on \protectedx{} reads (i.e., lock-free).  Epochs can solve the problem in constant time overhead, but with unbounded space. RCU ensures bounded space, but is blocking. We know of no other 
constant time, bounded space solution 
that does not require powerful primitives with operating system support. 
Based on our solution to the \imr{} problem we develop data structures for concurrent stacks and queues that improve on the time of peeking at the top of the stack or end of the queue.

\result{Stacks and Queues}{
  \label{result:stackqueue}
On $P$ processes, a collection of $M$ concurrent stacks or queues with a maximum of $N$ word-sized\footnote{Note the word could be a pointer to a much larger item, but we are not including the space of the larger item, just the pointer.}  items total
   across them can be supported with:
  \begin{enumerate}
    \item lock-free updates, and $O(1)$ time peeks,
    \item $O(M) + 2N + O(P^2)$ space, 
    \item single-word (at least pointer-width) read, write and CAS.
    \end{enumerate}
  }
  
The best previous implementations of concurrent stacks and queues either
require unbounded counters with double-word-width CAS~\cite{MS95,Tr86},
take unbounded time on peeks~\cite{michael2004hazard},
or require $O(M P^5 + N P^3)$ space and double-word-width LL/SC~\cite{aghazadeh2016tag}.


Next, we describe an implementation of reference counting for garbage collection.  Here, the resource is the reference count.  Copying a reference (pointer) to an object involves reading the reference and then incrementing the count on the object.  Destructing the
reference decrements the count on the object and collects it if the count goes to zero.  When the reference is kept in a shared mutable location, a read-destruct race can occur where a read and then increment is split by an overwrite, decrement and collect.  
To protect against this, we present the following result.

\result{Reference Counting}{
\label{result:refcount}
On $P$ processes, any number of reference counted objects with references stored in shared mutable locations
can be implemented safely with:
  \begin{enumerate}
   \item references as just pointers (i.e., memory addresses),
   \item $O(1)$ expected time for reading, copying and overwriting references,
   \item $O(P^2)$ space overhead and $O(P^2)$ \delayed{} decrements,
   \item single-word (at least pointer-width) read, write, CAS, FAS, and FAA.
  \end{enumerate}
}

We say a \emph{decrement is \delayed{}} if a reference has been overwritten or otherwise deleted, but the count on its object has not yet been decremented.
Previous approaches to protect against the read-destruct race in reference counting are either only lock-free~\cite{Valois95,MS95,DMMS02,Herlihy05,williams2012book}, wait-free with $O(P)$ time~\cite{Sundell05} per operation, require unbounded decrement-delay~\cite{McKenney07}, or use unbounded sequence numbers and double-word fetch-and-add~\cite{lee2010fast,plyukhin2015}, which is not available on modern machines. 
We note that although we are able to improve previous results for memory reclamation, stacks, and queues by simply applying our constant time acquire, for reference counting we have to change previous algorithms.   In particular, we cannot use sticky counters~\cite{Valois95,MS95,Herlihy05} since they need a lock-free CAS-loop instead of a single FAA for updating the reference count.  Our implementation avoids sticky counters by allowing multiple retires of the same resource. This was not allowed in previous interfaces.

Next, we generalize the approach used in reference counting collection to support objects that are managed with ownership---i.e., every object (resource) has a single owner, but objects can be copied or moved to another owner, or destructed when the object is dropped or the owner terminates.  This generalization is particularly useful in programming languages, such as C++ and Rust, that manage all objects with ownership and automatically insert move, copy and destruct methods into code at the appropriate locations (e.g., assignments, function calls, and end of scope)~\cite{Stroustrup2001}.  
The race that we want to protect against is between the copy and the destruct.   We assume that the copies and destructs are
safe if there are no such races, which we refer to as \emph{race-free safe} (Section~\ref{sec:copy_destruct}).
We say a \emph{destruction is delayed} between when the ownership of an object is released, i.e., ready to be destructed,
and when it is actually destructed.    We present the following result.

\result{Copy and Destruct}{
  \label{result:copydestruct}
  For race-free safe methods, the copy, move and destruct methods can be applied safely on shared mutable
  locations with
  \begin{enumerate}
    \item $O(1)$ expected time overhead per operation,
    \item $O(P^2)$ total space overhead, $O(P^2)$ delayed destructions,
    \item single-word (at least pointer-width) read, write, CAS, and FAS.
    \end{enumerate}
  } 
As far as we know, we are the first to consider this general form of protection for read-destruct races.  Reference counting is a special case in which the copy increments the count and the destruct decrements it.   The approach can also, for example, be used to safely manage the widely used C++ standard template library (STL) vectors and strings, as well as any other data structure that requires a deep copy.

\paragraph{Implementation.}
In addition to the theoretical results mentioned above we have implemented most of the techniques in C++.  The code is, perhaps, surprisingly simple and elegant.  For the copy and destruct idea, we develop a \cfont{weak\_atomic} that can wrap any type and make it safe for copy and destruct if it satisfied the copy-safe assumptions (almost always true).  The \cfont{weak\_atomic} can be wrapped around C++ shared 
pointers to give the equivalent of atomic shared pointers, but with constant time instead of using locks or being just lock free. 
In Section~\ref{sec:experiments} we describe some simple benchmarking that shows that \cfont{weak\_atomic} wrapped around C++ shared pointers performs well in practice. 

      \paragraph{Model and Assumptions.}
We assume the standard concurrent shared memory model with $P$ asynchronous processes and sequential consistency~\cite{HS08}.  Appropriate fences are needed for weaker memory models, and are included in our implementation.  We use the standard definitions of wait-free, lock-free and linearizability, invocation, and response~\cite{HS08}.  Throughout, when we talk about the \emph{time} of an operation, we mean the number of instructions (both local and shared) performed by that operation before it completes.
By \emph{pointer-width}, we mean a word that is just big enough for a pointer
---i.e., not even an extra bit hidden somewhere.  By space we mean number of words including both shared and local memory.
The ``expected time'' bounds we give are purely due to hashing.
Beyond reads and writes, we consider three other atomic read-modify-write primitives: \cfont{compare\_and\_swap} (CAS), \cfont{fetch\_and\_store} (FAS), and \cfont{fetch\_and\_add} (FAA).  All three instructions are supported by modern processors.

      \section{\acqret{}}
\label{sec:acquire_retire}

The \acqret{} interface can be used to efficiently protect against
read-destruct races and, as we will see in the following sections, can
solve a handful of important problems.  The interface consists of four
operations \acquire{}, \release{}, \retire{}, and \ejectx{}. The \acquire{} and \release{} operations are used to
protect and unprotect resources.  The \retire{} and \ejectx{}
operations are used to indicate when a resource is ready to destruct, and
when it is safe to destruct, respectively.

We assume that resources are accessed via a single-word \emph{handle}.
In the uses in this paper the handle is always a memory pointer, but
in general it need not be---e.g., it could be a key into a hash table,
or index into an array.
In the following discussion we say an
operation $x$ is \emph{linked} to another operation $y$, if $x$ maps
to just $y$, but there could be many operations that link to $y$. We
say $x$ and $y$ are paired, if they each are only matched to the other.
We now formally define the interface.






\begin{definition}[Acquire-Retire Specification]
\label{def:acquire_retire}
The acquire-retire interface supports the following four operations,
where \cfont{T} is the the type of a resource handle.
\begin{itemize}
  \item \cfont{T acquire(T* ptr, int i)}: returns the resource handle in the 
    location pointed to by \cfont{ptr} and ``protects'' it.
  \item \cfont{void release(int i)}: paired with the previous
    \cfont{acquire(*, i)} on the same process. Releases the
    resource protected by that \acquire{}.
  \item \cfont{void retire(T h)}: used after an atomic update on a
    location that overwrites a resource handle \cfont{h}.   
    The update is paired with the retire.
  \item \cfont{T eject()}: it is either paired with a previous
    \cfont{retire(h)} and returns \cfont{h}, or not and returns $\bot$.
  \end{itemize}
  We say that an
  \cfont{acquire(p, i)} links to the next update on location \cfont{p}
  (if any).  This update is then paired with a \cfont{retire} (if any) which is
  paired with an \cfont{eject} (if any).  By transitivity, every
  \cfont{acquire} is linked to at most one \cfont{retire}, and 
  the resource handle returned by the \cfont{acquire} is equal to the handle passed to its linked \cfont{retire}.
  The interface \textbf{guarantees} that for any \cfont{acquire},
  if it is linked to an \cfont{eject}, then 
  the \cfont{release} paired with the \cfont{acquire} must have happened before
  the \cfont{eject}.
\end{definition}

We note that the guarantee captures our intuition of  what the
interface is supposed to protect against.   In particular, it ensures
that any destruct of a resource placed after the \cfont{eject} will happen
after all processes \cfont{release} that resource.


Our interface allows multiple concurrent \cfont{retire}s of the same
handle, which leads to a subtle
point: every \acquire{} can be linked to (at most) one \retire{}, but
a \cfont{retire(h)} has the possibility of being paired with any
future \cfont{eject}, many which could return the same handle, i.e.,
one cannot tell by the handle returned by \cfont{eject} which ones are
matched.  Our guarantee is simply that there exists a matching of
\retire{}s to \ejectx{}s that satisfies the condition.

In most use cases, each process only needs to \protectx{} a single resource at a
time, so we omit the parameter \var{i} in \acquire{} and \release{}.
We assume that all four operations are atomic (i.e. linearizable).


\subsection{Related Interfaces}
\label{sec:related}


The acquire-retire interface is similar to the interfaces for hazard-pointers~\cite{michael2004hazard} and pass-the-buck~\cite{Herlihy05}.  A key difference is that those interfaces have a non-atomic acquire.  This means that after their ``weak'' acquire, they have to check the location to ensure the value has not changed in the meantime.  If it has, they have to try again.  For this reason a ``strong'' acquire, which we support, would be lock-free but not wait-free in their interface.  Another important difference is that our interface allows multiple concurrent retires on the same handle, but since theirs was designed specifically for memory reclamation, theirs insists that a handle is ejected before it can be retired a second time.  Allowing concurrent retires enables us to implement a reference counting collector that uses \cfont{fetch\_and\_add} instead of \cfont{compare\_and\_swap} for incrementing and decrementing counters.  It also allows us to define a copy-destruct interface that permits multiple destructs (e.g. an object that requires two destructs to kill it).  The interfaces also have some minor less important differences---e.g., in pass-the-buck, the eject is part of the retire.

Read Copy Update (RCU)~\cite{mckenney2008rcu} and epoch-based reclamation~\cite{fraser2004practical} also serve a similar purpose and have a somewhat similar interface.  Instead of acquiring and releasing individual resources, however, they \cfont{read\_lock} and \cfont{read\_unlock} regions that are not paired with any particular resource.  They guarantee that everything that is retired, by any process, during the protected (locked) region will be ejected after the region is finished.  In some cases, this can make protection easier.  On the other hand, it means a retired resource cannot be destructed until all protected regions that overlap the retire finish.  Most of the regions could have nothing to do with the particular retire.  This implies that the memory required by RCU and epochs can be unbounded.  The linux implementation of RCU mitigates this problem by disabling interrupts during a locked region and asking the user to ensure that they hold the read lock very briefly.  Brown~\cite{brown2015reclaim} gives bounds for a similar method.  These approaches require OS support.  Variants of epochs based on intervals~\cite{ramalhete2017brief,wen2018interval} can be used to bound the memory, but are not wait-free, and the memory bound is very large: $O(P M)$ where $M$ is the most memory that was live (allocated but not retired) at any time.

Ben-David et. al~\cite{ben2019vm} describe an interface similar to acquire-retire, and use it for multi-versioning.  It has an acquire, release, and set, where the set is equivalent to a CAS followed by a \cfont{retire}, and the \cfont{release} includes the \cfont{eject}.  They say an implementation of the interface is \emph{precise} if the last release that holds a resource (version) ejects the resource.  They describe a precise data structure that has constant time acquire, and $O(P)$ time release.  It requires a two-word wide CAS and uses unbounded counters.  Our implementation of acquire-retire is not precise, but we get $O(1)$ time release, and without two-word CAS or unbounded counters.  We conjecture it is not possible to be both precise, and $O(1)$ time for all operations of the acquire-retire interface.

\subsection{Implementing the \acqret{} Interface}
\label{sec:acqret-alg}
  \begin{figure}[!t!h]
  \caption{Implementing \acqret{}  for process $p_{pid}$}
  \begin{lstlisting}[linewidth=\columnwidth, xleftmargin=5.0ex, numbers=left]
shared variables:
  Destination<|T|> A[P][c];
local variables:
  list<|T|> rlist;
  list<|T|> flist;

T acquire(T* ptr, int i) {
  A[pid][i].swcopy(ptr);
  return A[pid][i].read(); }
void release(int i){A[pid][i].write(empty);}
void retire(T t) { rlist.add(t); }

optional<T> eject() {
  perform steps towards ejectAll(rlist); 
  if(!flist.empty())
    return flist.pop(); }

void ejectAll(list<|T|> rl) {
  list<|T|> plist = empty;
  for (int i = 0; i < P; i++)
    for (int j = 0; j < c; j++)
      plist.add(A[i][j].read());
  flist.add(rl \ plist);
  rlist.remove(rl \ plist); }
  \end{lstlisting}
  \label{alg:acqret}
  \end{figure}


Lock-free \acquire{} with constant time \release{}, \retire{}, and expected constant time \ejectx{}
can be implemented with techniques similar to the ones used in Hazard
Pointers~\cite{michael2004hazard}. To get \acquire{} down to constant time, we leverage 
a recently proposed primitive called \swcopy{}~\cite{BW19}.  The \swcopy{} atomically copies from one location to
another location, but requires that the destination location is only written to by a
single process. This primitive can be implemented in constant time with $O(P^2)$ space overhead~\cite{BW19}.

We begin by implementing the lock-free version.
As with Hazard Pointers, for a process to protect $c$ resources, it
owns $c$ slots in a shared \emph{announcement array} $A$.  The total
number of slots in $A$ is therefore $cP$.  A process uses
its $c$ slots to protect resources by placing their handles in these slots.

A lock-free \acquire(\var{T* ptr}, i) begins by reading \var{*ptr} and writing
the result in its \var{i}th announcement slot. To check if the announcement
happened ``in time'', it reads \var{*ptr} again to see if it has changed.
If it's still the same, then the announced handle has been 
successfully \protectedx{}, and can be returned. Otherwise, the \acquire{}
has to restart from the beginning.

A \release(i) operation unprotects by simply clearing the process's \var{i}th announcement location, and 
\retire(\var{T x}) simply adds $x$ to a process local retired list called \var{rlist}. 
To determine which handles are safe
to eject, we implement an \ejectall{}(\var{rl}) operation which first loops 
through $A$ and makes a list of all the handles that it sees.
We call this list of handles \var{plist} for ``protected list''.
If a handle is seen multiple times
in $A$, then it will also appear that many times in \var{plist}.
Next, \ejectall{} computes a multi-set difference between \var{rl} and
\var{plist} (denoted \var{rl \textbackslash{} plist}). 
The result of this multi-set difference are handles that can be safely ejected without
violating the specifications from Definition \ref{def:acquire_retire}.
It's important that we keep track of multiplicity and perform multi-set
difference because in the case where there are multiple \retire{}s of the
same handle, each occurrence of this handle in the announcement array might
be linked to a different \retire{}. So if a handle appears in \var{rlist} 
$s$ times and the announcement array $t$ times, it is safe to eject only 
$s-t$ copies of this handle.

There are two ways of performing the multi-set difference. The most 
general method is to use a hash table and this would result in a $O(N + cP)$
\emph{expected} time \ejectall{}(\var{rl}), where $N$ is the size of \var{rl}.
If \ejectall{}(\var{rlist}) is run once every $O(cP)$ \retire{}s, then we can guarantee that
\var{rlist} does not become too big. This is because only $cP$ handles
can be \protectedx{} at any time, so if \var{rl} has more
than $2cP$ handles, then \ejectall{}(\var{rl}) will add at least $cP$ of them to the free list.
The size of \var{rlist} is always upper bounded by $O(cP)$.

An \ejectx{} is essentially a deamortized version of \ejectall{}. Every time it is called, it performs a small (expected) constant number of steps towards \ejectall{}(\var{rlist}). When \ejectall{} returns a list of handles, they get stored in a local free list
to be returned one at a time by the following \ejectx{}s.
If \ejectx{} is called after every \retire{}, then an entire \ejectall{}() will complete every $O(cP)$ calls to \ejectx{}. To ensure that this is the case, \ejectx{} 
does not breakup hash table operations. Thus, it takes expected constant time.


Finally, we turn our attention to making \acquire{} constant time. This can be done by making the read of \var{*ptr} and write to the announcement array appear to happen atomically. This is exactly the functionality 
provided by the Single-Writer Copy (\var{swcopy}) primitive \cite{BW19}.
To use this primitive, we would have to replace each element of the announcement array with a \var{Destination} object. These \var{Destination} objects support read, write and \swcopy{}(\var{T* src}), and they are single-writer, which means that only one process is allow to write
and copy into each object.
The \var{swcopy} primitive takes a pointer to an arbitrary memory location and appears to atomically copy
the word from that memory location into the \var{Destination} object.
Blelloch and Wei \cite{BW19} present an implementation of $M$ 
\var{Destination} objects using $O(M+P^2)$ space such that \var{read},
\var{write} and \var{sw-copy} all take constant time. In our implementation, we use $O(cP)$ such objects for the announcement array.

Pseudo-code for this implementation appears in Figure \ref{alg:acqret}.




\paragraph{Proof outline of Result~\ref{result:acqret}.}


We need to prove that for an acquire, the eject that it links to (if
any) happens after the release it is paired with.  We consider an
\cfont{acquire} returning \cfont{h}, the \cfont{release} it is paired
with, the update it is linked with, the \cfont{retire(h)} that the
update is paired with, and the \cfont{eject} that the \cfont{retire} is paired
with.  Since \cfont{swcopy} is atomic, the \acquire{} is linearized
at the copy point.  After this point and before the \cfont{release},
the announcement array will contain the protected handle \cfont{h}.
The next update in linear order on the same location is linked to the
acquire, and this is in turn possibly paired with a \cfont{retire(h)}.
If the \cfont{retire} happens after the \cfont{release}, then clearly any paired
\cfont{eject} will happen after the \cfont{release}, so we assume the
  \cfont{retire(h)} happens before the \cfont{release}.  Between the \cfont{retire} and
the \cfont{release} the announcement array contains a copy of \cfont{h} for
the specific \cfont{acquire} (it could also hold other copies).  Because of
the multi-set difference applied between the retired list and
announcement array, every \cfont{h} in the announcement array can be
paired with up to one element in the retired list of a process,
and that one will not be ejected.  Therefore, by the pigeonhole
principle, any that are ejected are no longer protected.  Hence, the
\cfont{eject} linked to an \cfont{acquire} can only happen after the paired \cfont{release} that removes \cfont{h}
from the announcement array.

The time for \cfont{acquire}, \cfont{release}, and \cfont{retire} are
constant. The time for \cfont{eject} is constant in expectation due to the process local hash table that is used for set difference.
If \ejectx{} is called after each \retire{}, the space is bounded by $O(cP^2)$ since each process
requires at most $O(cP)$ space. We say that a retire-eject pair is  \emph{\delayed{}} between the \cfont{retire} and the \cfont{eject}. The number of delayed
retire-eject pairs is bounded by $O(cP^2)$ since each process can have
at most $O(cP)$ in its retired list, $O(cP)$ in a partially
completed \cfont{ejectAll} and, $O(cP)$ in the results from the
previous \cfont{ejectAll} that have not been ejected yet. The implementation only uses atomic single word read, write and CAS.

      \section{Memory Reclamation}
\label{sec:reclamation}

The memory reclamation problem is the special case of read-destruct races where the resource is a pointer to a block of memory, and the destruct is a call to the free of that block---i.e., to return the block to a pool for later reuse.  In this context, the race is sometimes called a read-reclaim race~\cite{hart2007performance} and has been widely studied~\cite{michael2004hazard,brown2015reclaim,Herlihy05,mckenney2008rcu, fraser2004practical,ramalhete2017brief,wen2018interval}.  We note that in this special case, there can only be one concurrent destruct on a resource (block) otherwise we would double free a block.  

 \begin{figure}
 \begin{lstlisting}[linewidth=.99\columnwidth, xleftmargin=5.0ex, numbers=left]
template<|typename R, typename T, typename F|>
R protected_read(T** location, F f) {
  T* ptr = acquire(location);
  R result = f(ptr);
  release();
  return result; }

template<|typename T|>
void safe_free(T* ptr) {
  retire(ptr);
  optional<|T*|> ptr2 = eject();
  if (ptr2.has_value()) free(ptr2.value()); }
\end{lstlisting}
   \caption{Protected-block memory reclamation interface and code.  The \cfont{location}
       is a pointer to a pointer, and the acquire atomically reads from \cfont{location} and
       acquires the pointer that is read.  The user function $f$ takes a pointer (\cfont{T*}) as an argument and can return any type (\cfont{R}).}
  \label{figure:reclamation}
\end{figure}

Within the context of memory reclamation there are two classes of interfaces that both protect read-reclaim races.  In the first, supported by hazard pointers and pass-the-buck, individual memory blocks are protected.  We refer to this as protected-block memory reclamation.  In the second, which includes RCU and epochs, regions of code are protected along with all the blocks that are read in the region.  This can be referred to as protected-region memory reclamation.  There is a tradeoff between the interfaces.  The first can support bounds on the number of resources used even if threads stall or fail, while the second can more easily protects a large collection of objects.  Here, we are interested in the protected-block version.  In this case making memory access safe for read-reclaim races just involves wrapping the use of a memory block in an acquire-release on the block, and splitting the reclamation of memory into a retire and then an eject that frees the pointer at some later point when it is safe.   The interface and code is shown in Figure~\ref{figure:reclamation}.

We say that a memory block pointed to by $p$ is \emph{delayed} if \cfont{safe\_free}$(p)$ has been called but the corresponding
\cfont{free}$(p)$ has not yet been called.
Proper usage of protected-block memory allocation requires that while a pointer is delayed, no \cfont{protected\_read(loc)} can use that pointer (i.e., it cannot be stored in \cfont{loc}), and no \cfont{safe\_free} can be called on the pointer.  The first corresponds to a use after free and the second to a double free.

\paragraph{Proof outline of Result~\ref{result:reclamation}.}
We first consider safety---i.e., with proper usage, a block is never freed during a protected read.  This follows
from the proper usage requirement that if the block is protected after it is retired, the protected regions must have been acquired before the retire.   Due to the semantics of acquire-retire, the eject cannot happen until after these protected regions are all released, and hence the free will happen after all releases and is safe.

We now consider the five properties.  The first two about constant time overhead follow directly from the acquire-retire results. 
The $O(cP^2)$ space (3) also follow from the acquire-retire results.  The number of delayed blocks is bounded by the $O(cP^2)$  delayed retires given by the acquire-retire result (4).  The implementation uses only the primitives needed by acquire-retire (5).

 \begin{figure}
 \begin{lstlisting}[linewidth=.99\columnwidth, xleftmargin=5.0ex, numbers=left]
template<|typename T|>
struct stack { 
private:
  struct Node {T value; struct Node* next;};
  struct Node *Head;
public:
  stack() : Head(nullptr) {}
  
  optional<T> peek() {
    return protected_read(&Head, [] (Node* p) {
       if (p == nullptr) return optional<T>();
       else return optional<T>(p->value);});}

  void push(T v) {
    Node* a = allocate(sizeof(Node));
    a->value = v;
    auto try = [&] (Node* p) {
      a->next = p;
      return CAS(&Head, p, a);};
    while (!protected_read(&Head,try)) {};}

  optional<|T|> pop() {
    Node* r;
    auto try = [&] (Node* p) {
      r = p;
      if (p == nullptr) return true;
      return CAS(&Head, p, p->next);};
    while (!protected_read(&Head, try)) {};
    if (r == nullptr) return optional<T>();
    T a = r->value;
    safe_free(r);
    return optional<T>(a);} };
\end{lstlisting}
   \caption{A concurrent stack with peek, push and pop.} 
  \label{fig:stack}
\end{figure}

\paragraph{Stacks and Queues.}   Protected-block memory reclamation can be used to implement lock-free stacks and queues with a single word CAS and without unbounded tags~\cite{michael2004hazard,Herlihy05}.    The standard ABA problem is avoided
since a link (memory block) in the structure cannot be recycled while someone is doing a protected access on its pointer.   However, by doing this, even just peeking at the top of a stack or the front of a queue requires an acquire.   Therefore the Hazard pointers and pass-the-buck only support lock-free peeks.    By using our acquire-retire algorithm, we get constant time peeks, while preserving the time for \cfont{push} and the time for \cfont{pop} in expectation.   The code for \cfont{peek}, \cfont{push},  and \cfont{pop} for stacks is shown in Figure~\ref{fig:stack}.   The \cfont{push} and \cfont{pop} are effectively the same as previous results~\cite{michael2004hazard,Herlihy05}.   
The code for \cfont{enqueue} and \cfont{dequeue} on queues is again the same as the previous results and the peek is easy to implement (not shown).

\paragraph{Proof outline of Result~\ref{result:stackqueue}.}
The correctness and lock-free updates follow from previous work~\cite{michael2004hazard,Herlihy05}.    The time bounds for peek (1) follow from the fact that the \cfont{protected\_read} is $O(1)$ time.   The space includes a constant number of words
for each stack or queue ($O(M)$), plus two pointers per node in the linked lists for the value, and the next pointer  ($2N$) , plus the space overhead of protected-block memory reclamation ($O(P^2)$).   The total is $O(M) + 2N + O(P^2)$ as claimed (2).   We note that for stacks the constant in the $O(M)$ is one.
The implementation only uses single word atomics (3).

      \section{Reference Counting}
\label{sec:ref_counting}

Reference counting (RC) garbage collectors~\cite{Collins60} keep a
count associated with every memory block managed by the collector,
indicating how many pointers reference the block.  Whenever a pointer
to the block is copied, this count is incremented, and whenever a
pointer is overwritten or destructed, the count is decremented.  If
the count goes to zero, the block itself is destructed, which in
addition to freeing memory will destruct any pointers within the
block, possibly causing more blocks to be destructed.  Reference
counting is widely used both in collected languages (e.g., python and
swift) and in non-collected languages managed by ownership (e.g., C++
with \cfont{std::shared\_ptr} and Rust with RC and ARC pointers).
With ownership,
the counters can be automatically incremented and decremented by copy
and destruct methods.

Compared to other garbage collection schemes, such as stop-and-copy
and mark-and-sweep, reference counting has the advantage that it more
aggressively collects garbage, often as soon as it becomes
unreachable.  However, it has the disadvantage that it cannot collect
cycles in the memory graph.  There are techniques to alleviate this
problem, such as weak pointers or using separate collectors for
cycles.  

In this paper, we are interested in using RC in the concurrent
setting.  Reference counters can be incremented and decremented
concurrently with an atomic FAA instruction.  If the references are
thread private or immutable, this is all that is needed.  However, if
the references are held in locations that can be simultaneously read by
one thread and updated by another, which is true in just about any
lock-free data structure, then there can be a read-destruct race.  In
particular the read will read the location, increment the count, and
return the location, while an update will overwrite the location and
then decrement the count of the old value, possibly collecting it if
the count goes to zero.  If an update, decrement and collect splits the read and increment,
then the memory block will be collected before it is incremented and
returned, hence incrementing a collected counter and returning a
pointer to garbage.


\paragraph{Related work.}
The read-destruct problem for reference counts is well understood.
Valois, Michael and Scott~\cite{Valois95,MS95} first developed a
lock-free approach to prevent it.  Their approach, however, can
increment the counter of freed memory and requires a CAS to update
the counter, which could fail and need to be repeatedly retried.
Detlefs et al.~\cite{DMMS02} describe lock-free method that avoids
these two issues, but it requires a DCAS (two word CAS), which is not
supported by any machine.  Herlihy et al.~\cite{Herlihy05} are able
to remove the DCAS assumption, leaving us with a lock-free solution to
the problem with just single-word CAS, but it still requires a CAS loop
instead of a FAA due to the need of a sticky counter.  Most
importantly, all these solutions are lock-free but not wait-free.  In
particular, a thread trying to read a pointer could retry indefinitely
as other threads update or copy the pointer.


Sundell developed a wait-free solution~\cite{Sundell05}.  However,
like the Valois-Michael-Scott method, it can increment freed memory,
making it inappropriate in many practical situations.  Also, and
perhaps more critically, the approach requires $O(P)$ time to retire
each location, which is expensive.  RCU and epochs can also be
used, but with no bound on memory usage since GC can be delayed
arbitrarily.
From a practical standpoint, the C++ STL library supports atomic
shared pointers, which are reference counted pointers that are safe
for concurrency.  The standard implementations used in GCC and LLVM
are based on locks.  However, there is at least one library that support
a lock-free solution~\cite{just19}.  It is based on split reference
counters~\cite{williams2012book}.  It
requires double word CAS and is lock-free, but not wait-free.
A similar technique was independently developed by Lee~\cite{lee2010fast}
and generalized by Plyukhin~\cite{plyukhin2015}. Their version requires 
atomic double-word FAA on a location containing both a pointer and an 
unbounded sequence number. Unlike double-word CAS, double-word FAA is not 
supported by modern machines.

As far as we know, no previous work simultaneously supports constant time 
access, has bounded memory overhead, and uses only instructions supported 
by hardware. 
Here, we support constant time operations on mutable
pointers to reference counted objects.  We do this by protecting the
read and increment of a reference pointer with an acquire/release, and
using a retire instead of decrement when overwriting an old pointer.
We ensure that no more than $P$ locations per process ($P^2$ total)
have been retired but not decremented.

\begin{figure*}
\begin{minipage}[t]{.49\textwidth}
\begin{lstlisting}[linewidth=.99\textwidth, xleftmargin=5.0ex, numbers=left]
template <|typename T|> 
struct ref_ptr {
  ref_ptr() : p(nullptr) {}
  ref_ptr(T* p) : p(p) {
    p->add_counter(1); }

  // copy pointer from b into 
  // new ref_ptr
  ref_ptr(const ref_ptr& b) {  @\label{line:refcopy}@
    p = acquire(&(b.p)); 
    if (p) p->add_counter(1); 
    release();}

  // update this ref_ptr with 
  // new ref_ptr b
  void update(ref_ptr b) { @\label{line:refassign}@
    T* old = fetch_and_store(p, b.p);
    retire(old);
    b.p = nullptr; @\label{line:refclear}@
    optional<|T*|> e = eject();
    if (e.has_value()) 
      decrement(e.value());}

  template<|typename R, typename F|> 
  R with_ptr(F f) { @\label{line:refwith}@
    R r = f(acquire(&p)); release();
    return r; }

  ~ref_ptr() { decrement(p); } @\label{line:refdestruct}@

private:
  T* p; 
  void decrement(T* o) { @\label{line:refdec}@
    if (o != nullptr && 
        o->add_counter(-1) == 1) 
      delete o;} };
\end{lstlisting}
    \center (a) Reference counted pointer
\end{minipage}\hspace{.3in}
\begin{minipage}[t]{.49\textwidth}
\begin{lstlisting}[linewidth=.99\textwidth, xleftmargin=5.0ex, numbers=left]
struct node {
  int cnt;
  int val;
  ref_ptr<|node|> Left, Right;
  node(int v, 
       ref_ptr<|node|> L, 
       ref_ptr<|node|> R) :
    cnt(0), val(v), 
    Left(L), Right(R) {}
  int add_counter(int i) {
    return fetch_and_add(count, i);}
};

using np = ref_ptr<|node|>;

// create tree Tr with three nodes
np Tr(new node(5, 
        np(new node(3, nullptr, 
                       nullptr)),
        np(new node(7, nullptr, 
                       nullptr))));

// update Tr in another thread 
thread other([&] () {
  Tr.update(np(new node(11,nullptr,
                           nullptr)));});

// copy Tr into Tr2, races with 
// update on previous line
np Tr2(Tr);

// join thread
other.join();
\end{lstlisting}
  \center (b) An example (safe) application to trees
\end{minipage}
\caption{Reference counting garbage collection (in C++), and an example use for
  maintaining a binary tree.}
\label{fig:ref_count}
\end{figure*}

\paragraph{Our implementation.}
Figure~\ref{fig:ref_count}(a) gives a safe implementation of a
mutable reference to a counted object.  It
assumes the object has an \cfont{add\_counter} method that can be used
to atomically increment or decrement the counter.  The copy
constructor (line~\ref{line:refcopy}), copies a pointer by acquiring the pointer, incrementing the count of the object being pointed to, setting
\cfont{p} to the copied pointer, and then releasing the pointer.  The
update function (line~\ref{line:refassign}) uses a \cfont{fetch\_and\_store} to swap in the new
value and return the old value.  It then retires the old value and
runs an eject step.  The eject can return a previously retired
pointer.  If it does, then the pointer is decremented.
Line~\ref{line:refclear} is needed since C++ will insert a destructor
for $\cfont{b}$ on exit of \cfont{update} and we do not want this to
decrement the counter for what is now stored in \cfont{p}.

The \cfont{with\_ptr} function (line~\ref{line:refwith}) takes a
function as argument, and applies the function to the raw object pointer.  As with the copy, this is wrapped in an acquire and
release to protect the access.  The
\cfont{decrement}$(o)$ (line~\ref{line:refdec}) attempts to garbage collect $o$ by
decrementing its count and checking if it went to zero (the
result returned by \cfont{add\_counter} is the old value).  If it went
to zero, it deletes $o$.  The destructor for $o$ can then destruct
other reference counted pointers, potentially recursively collecting
more objects.  Note that the destructor (line~\ref{line:refdestruct}) calls \cfont{collect}
immediately instead of calling the delayed \cfont{retire}.  This is
because in a proper program, when the pointer itself is being
destructed, no one else will hold a copy of the pointer so a
read-destruct race cannot occur.

\paragraph{Example.}
Figure~\ref{fig:ref_count}(b) gives an example use of \cfont{ref\_ptr}
for binary trees.  A tree node consists of the reference count, a
value, and a left and right child.  The constructor sets the counter
to 0, and sets the value and children appropriately.  The \cfont{add\_counter}
atomically updates the count.  The example below the node definition
creates a tree with three nodes.  It then concurrently updates and
copies the tree.  This would normally create a read-destruct race and
would be unsafe (e.g., using STL \cfont{shared\_ptr}).  It is,
however, safe with our version.  After the join point, and depending
on the outcome of the race, \cfont{Tr2} will either contain the
original tree or the new singleton tree.  If it contains the new tree,
then the original was retired by the update and the reference count on
the root of the new tree is 2.


\paragraph{Proof outline of Result~\ref{result:refcount}.}
We first consider safety.  From Definition~\ref{def:acquire_retire} of
the acquire-retire interface, every acquire is linked to at most one
retire. This happens through the update that overwrote the acquired handle.
Furthermore, the linked \cfont{eject} will happen after the release paired
with the acquire.    Since the \cfont{decrement} happens after the
\cfont{eject}, the \cfont{decrement} must happen after the protected
region of the acquire.  Therefore, during any copy of a pointer, or
protected read using \cfont{with\_ptr}, a decrement due to the
next update of the location cannot happen until after the
protected region.   Hence, the counter will never go to zero and be collected
during the protected region.

We now consider the four properties from Result~\ref{result:refcount}.  (1) References are just pointers, as
claimed.  The $O(1)$ expected time (2) and $O(P^2)$ space (3) follow
directly from the acquire-retire results.  Note that we cannot use the
worst-case version since we have multiple retires on a single
resource.  Similarly, number of delayed decrements is at most $O(P^2)$ (3) because there are at most $O(P^2)$ delayed retires. The implementation uses the primitives needed by
acquire-retire and additionally a FAA for incrementing and decrementing
the reference count (4).

\paragraph{Generalization.}
In the next section we generalize the idea we used in reference
counters to a broad set of resources that are based on copying and
destructing.  The generalized approach actually has a practical
advantage over what we described even in the context of reference
counting in that it more cleanly separates private or immutable
pointers to reference counted objects from shared ones.  This allows
it to avoid an acquire/release on pointers that are private or
immutable.

      \section{Copy and Destruct}
\label{sec:copy_destruct}

Languages based on the resource allocation is initialization (RAII) paradigm (e.g., C++, Ada, and Rust) manage resources (objects and values) by ownership~\cite{Cpp15,Jung17}.  In this setting, objects have one owner that is responsible for destructing the object when it is overwritten or goes out of scope.   This typically uses a destruct method, which can be defaulted or user defined.  Most objects also have copy and move methods that are used to make a copy of the object or move the object to a new owner.  The compiler inserts calls to the copy, move, and destruct methods throughout the code where needed to ensure single ownership and proper destruction, e.g. at function calls, assignments, the end of lexical blocks, and exception points.\footnote{We note that C++ copies by default when passing by value or assigning, while Rust moves by default.  In both languages the other choice can be made by an explicit move or copy (called clone), respectively.}  Ownership is particularly powerful in languages that are not garbage collected since, if used properly, it avoids any memory leaks or other memory problems, even in the presence of exceptions.  Typical examples of objects managed using non-trivial copy, move and destruct methods are strings or vectors in the C++ standard template library (STL).  For strings the copy allocates a new block of memory, copies the characters and returns a new header with the length and other information, along with a pointer to the new memory.  The move just moves the header, clearing the old location. The destruct will free the memory block, and clear the header.  Another example is shared reference counted pointers as discussed in the previous section, where the copy increments the counter, and the destruct decrements the counter and deletes the object pointed to if the counter goes to zero.  The approach is also often applied to nested objects such that the copy does a deep copy of the object.  Ownership can also be used for locks (the destruct releases the lock), and file pointers (the destruct closes the file), but in these cases the copy is often disabled.

Here, we generalize the approach used for reference counting to other objects managed by ownership.  As with reference counts, we are concerned with read-destruct races than can occur when an object is stored in a shared mutable location, where the race is between a load operation (consisting of a read of the location followed by applying the copy on the object), and a store operation (consisting of swapping a new value into the location, and then destructing the old value).  As previously mentioned, the problem occurs when the store fully splits the read and copy.

In the following discussion, we assume that all objects have a method $\cpy(o)$ that makes a copy of $o$, returning it, and a method $\destr(o)$ that destructs the object $o$, returning nothing.  We assume objects of type $\typ$ have an empty instance, denoted as $\emp{\typ}$, and that $\destr(\emp{\typ})$  does nothing.
We consider the following interface for a mutable reference cell for storing objects of type $\typ$.

\begin{description}
\item[$\newr(o)$:] creates a cell initializing it with the object $o$
\item[$\loa()$:] reads the object stored in the cell, makes a copy, and returns it;
\item[$\str(o)$:] stores the object into the cell, calling \destr{} on the old object;
\item[$\mov()$:] writes $\emp{\typ}$ into the cell, and returns the old value, and
\item[$\del(o)$:] deletes the cell and destructs its contents.
\end{description}
We note this interface supports ``ownership'' since it copies the value on reading it (\loa), and
destructs the owned value when overwriting it (\str) or destructing the cell itself (\del).
Figure~\ref{mutablecell}(a) gives C++-like code that supports the interface based on $\cpy$ and $\destr$.  The $\newr$ is defined as the constructor in the fourth line, and the $\del$ as the destructor in the last line.
In the code, we are explicitly inserting the copy and destruct methods---C++ itself will put them in implicitly.

\begin{figure*}
  \small
\begin{minipage}[t]{.51\textwidth}
  \begin{lstlisting}[linewidth=.99\textwidth, xleftmargin=5.0ex, numbers=left]
template <|typename T|>
struct mutable_ref {
  T p; 
  mutable_ref(T o) : p(o) {}

  T load() {
    T o = copy(p);

    return o; }

  void store(T o) {



    destruct(fetch_and_store(p, o));}
     
  T move() {
    return fetch_and_store(p, empty);}
 
  ~mutable_ref() { destruct(p);} };
\end{lstlisting}
\center (a) Unsafe version 
\end{minipage}\hspace{.3in}
\begin{minipage}[t]{.47\textwidth}
\begin{lstlisting}[linewidth=.99\textwidth, xleftmargin=5.0ex, numbers=left]
template <|typename T|>
struct mutable_ref {
  T p;
  mutable_ref(T o) : p(o) {}

  T load() {
    T o = copy(acquire(&p));@\label{line:copycopy}@
    release();@\label{line:copyrelease}@
    return o; }

  void store(T o) {
    optional<|T|> e = eject(); 
    if (e.has_value()) 
      destruct(e.value()); 
    retire(fetch_and_store(p, o));}
     
  T exchange(T o) {
    return fetch_and_store(p, o);}
 
  ~mutable_ref() { destruct(p);}};
\end{lstlisting}
\center (b) Safe version 
\end{minipage}
\caption{Code for supporting mutable reference cells with ownership.   Assumes C++ is not inserting its own copy constructor and destructor on the type \cfont{T}.   Our C++ implementation does not make this assumption.}
\label{mutablecell}
\end{figure*}

Our goal is to create a implementation of the interface that protects against a read-destruct race between the $\loa$ and the destruct in the $\str$.  To support this
efficiently we assume the type $\tau$ can be atomically read and swapped.  This is without loss of generality (although perhaps some loss of efficiency) since if the object is larger than an atomic word, it can be stored indirectly with a pointer to the object, i.e., it can be boxed.  

We require that that the supplied copy and destruct methods are safe when there are no read-destruct races.  In particular we say the copy and destruct are \emph{race-free safe} if they are safe when a copy and its linked destruct (if any) never overlap in time.  The destruct linked with a copy follows the definition of the acquire-retire interface.  In particular the copy of an object from a location links with the next update of the object in that location (if any), which is then paired with a destruct of the object. 
We note that copy and destruct methods are typically race-free safe.  This is because it is typically safe to run multiple $\cpy$ operations on the same source object concurrently since \cpy{} often just reads the shared source object, and writes to the local (unshared) copy that is returned.  This is true even if the \cpy{} is a deep copy.  In many cases there is only one destruct, so as long as it does not overlap with the copy it is safe.  For shared pointers, the \cpy{} updates a shared reference count, but this increment can be done atomically with a fetch-and-add.  There can also be overlapping destructs on the same count (since a destruct can be applied multiple times), but these are safe with a atomic decrement.  As long as copies do not overlap their linked destruct, the counter will only go to zero when no references are left and there are no ongoing copies.

Assuming the copy and destruct are race-free safe, our implementation can work as shown in Figure~\ref{mutablecell}(b).  We note that there are basically only three changes.  The first is to wrap the \cpy{} that is part of the load inside of an acquire release.  The second is to use retire instead of destruct in the store, and add an eject which does the destruct later when safe.  The third is to use atomic swaps instead of regular swaps for the store and move. It's possible to add a CAS to this interface and implement it similarly to store.  We note that the destructor for the reference itself can use a \destr{} instead of \cfont{retire} on $p$ since correct use of the reference (or any object) requires that all operations on it respond before invoking the destruct.

\paragraph{Proof outline of Result~\ref{result:copydestruct}.}
The correctness proof follows our other proofs.   In particular, if an acquire links to an eject, the eject will happen after the paired
release.   Therefore, the copy will be finished before the object is destructed.   Along with being race-free safe
this ensures correctness.
The bounds on time, space and delayed destructions also follow from acquire-retire interface.

\paragraph{Applications}
In our C++ implementation we have created a templated class:
\begin{lstlisting}[frame=none]
  template <|typename T|>
  struct weak_atomic; 
\end{lstlisting}
It supports protected copies and destruct if the methods are race-free safe.   For example, It can be used safely for
the C++ STL structures \cfont{shared\_ptr}, 
\cfont{vector}, and \cfont{string}.
All these have copy and destruct methods that are race-free safe.  The \cfont{weak\_atomic} interface supports a \cfont{load}, \cfont{store} interface similar to Figure~\ref{mutablecell}, which  is similar to the C++ \cfont{atomic} structure.  When used as \cfont{weak\_atomic<shared\_ptr>}, for example, it gives a safe type for reference counting that can be a plug in replacement for \cfont{atomic<shared\_ptr>}.    We use this in the experiments in the next section.

\section{Reference Counting Experiments}
\label{sec:experiments}

\begin{figure*}
    \includegraphics[width=\textwidth]{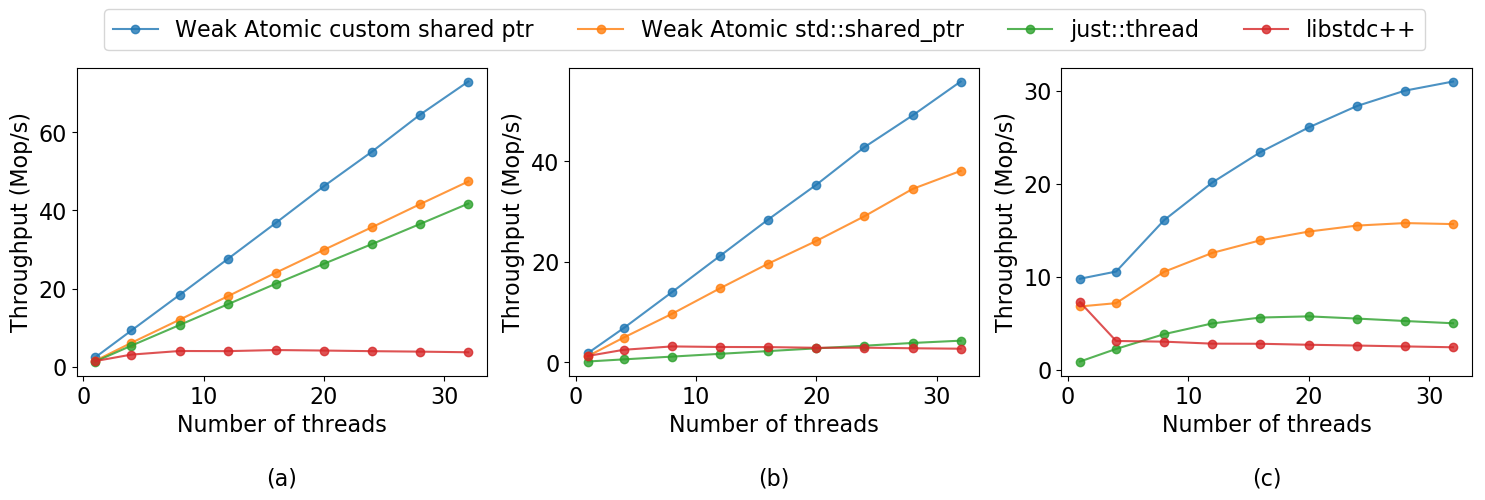}
    \caption{Comparison of different reference counting implementations under various workloads. Figures (a) and (b) are both low contention ($N=10M$), and figure (c) is high contention ($N=10$). The percentage of stores are $0\%$, $50\%$, and $10\%$ for Figures (a), (b), and (c) respectively. }
    \label{fig:exp}
\end{figure*}

In this section, we present some experimental results for our concurrent 
reference counted garbage collector presented in Section \ref{sec:ref_counting}.
These experiments are not meant to be a comprehensive evaluation of existing 
concurrent garbage collection techniques nor of our acquire-retire interface.
Instead, their purpose is just to show that our approach is 
light weight enough to be used in practice and that it scales well.

\textbf{Setup.} We ran our experiments on a 4-socket machine with 32 physical cores (AMD Operton 6278 processor), 2-way hyperthreading, 2.4 GHz, 6MB L3 cache, and 200 GB main memory. All our experiments were run up to 32 cores without hyperthreading and we interleaved
memory across sockets using \var{numactl -i all}. For scalable memory allocation
we used the jemalloc library \cite{jemalloc}. All our experiments were written in C++ and
compiled with g++ version 9.2.1 on optimization level \texttt{O3}.

\textbf{Workloads.} The workloads we ran involve loads and stores on an array of $N$ reference counted pointers, each pointing to a different object. This array is padded so that each pointer is on a different cache line.
A load involves reading the pointer and creating a new reference counted pointer
to that object. A store involves allocating a new object and changing a reference
counted pointer to point to that object.
The location of the load/store is picked uniformly randomly between $0$ to $N-1$, so smaller $N$ means more contention. We show results for $N=10$ (representing a highly contended workload) and $N=10M$ (representing a workload with almost no contention). Note that for $N=10M$, less than 1\% of the array fits in L3 cache.
Stores are performed with probability $p_s$, and we try three different settings:
load only ($p_s = 0$), load mostly ($p_s = 0.1$), and store heavy ($p_s = 0.5$). 
We run each experiment for 3 seconds and report the overall throughput of loads
and stores (averaged across 5 runs).

\textbf{Implementations.} We compare two implementations based on our approach with the \atomicsharedptr{} implementation in the GNU C++ library 
\cite{gnulib}. We also compare with a commercial implementation of \atomicsharedptr{} from Anthony William's \justthread{} library \cite{just19}. All four implementations provide a similar interface and solve the same problem.
The implementation in GNU is lock-based whereas \justthread{} is lock-free, and according to this post \cite{just19}, it uses
something similar to the split reference count technique described in \cite[Chapter~7.2.4]{williams2012book}.

For our implementations, we use the \cfont{weak\_atomic} class described in Section~\ref{sec:copy_destruct}, which can be wrapped around any type and make it safe for copy and destruct if it satisfied the copy-safe assumptions.  In one implementation, we wrap \cfont{weak\_atomic} around C++'s shared pointer type, \stdshared{} \cite{sharedptr}, to achieve the equivalent of \atomicsharedptr{}. We refer to this implementation as \weakatomicstd{}.  We also implement our own shared pointer type which is more efficient, but less general compare to \stdshared{} since it does not support weak pointers.  We refer to this implementation as \weakatomiccustom{}. In our implementation of \acqret{}, we applied a fast-path/slow-path optimization where the \acquire{} operation runs the lock-free version of the \acquire{} for a few tries before switching over to the wait-free version. 


\textbf{Analysis.} We selected three graphs from our benchmarks to show in Figure \ref{fig:exp}. Across all our benchmarks, both versions of our implementations seemed to scale reasonably and perform better than the two competitors. The \justthread{} library performed better whenever we lowered the contention level or lowered the frequency of stores. The load only workload in Figure (a) shows the closest \justthread{} gets to the throughput of our implementations. The single-core perform of the GNU implementation is actually very close to \weakatomicstd{} in all the workloads we tried, but it achieves relatively little speed up (less than 3x on 32 cores). This is because the GNU implementation shares a small number of global locks across all $N$ pointers.

In the low contention case (Figures \ref{fig:exp} (a) and \ref{fig:exp} (b)) where it is rare for two processes to access the same pointer, we observed that both our implementations scale nearly perfectly, achieving 30-31x speedup on 32 cores. The \justthread{} library also 
achieves 31x and 29x speedup on workloads (a) and (b) respectively. While \justthread{} gets a lot of self-speedup in the store-heavy workload (Figure (b)), its single-core performance is over 8x slower than ours. In the high contention case (Figure \ref{fig:exp} (c)), our implementations still manages to achieve modest scaling.

\section{Conclusion}
\label{sec:conclusion}

We have presented a constant-time per operation and bounded-space
interface for protecting against read-destruct races, and used it for
a handful of important applications.  While our specific results
are important on their own, we believe the framework is also
important.  By considering resources in general, our acquire-retire
interface generalizes previous interfaces, which focused mostly on the
memory-reclamation problem.  We allow, for example, multiple
retires/destructs on the same resource, and define a general
copy-destruct interface that covers a reasonably broad set of
applications.  The \cfont{weak\_atomic} class, based on
the copy-destruct interface, is able to implement a concurrent reference
counting collector in C++ as easily as
\cfont{weak\_atomic<shared\_ptr>}.  Furthermore our preliminary
experimental results indicate that it is fast.


\section{Acknowledgments}
We would like to thank Daniel Anderson for his C++ experitise and for improving our code.

\bibliographystyle{abbrv}
\bibliography{../../biblio}


\end{document}